\documentclass[twoside,showpacs,superscriptaddress,twocolumn,floatfix,a4paper,aps]{revtex4}

\usepackage{color}
\usepackage{graphicx}
\usepackage[utf8x]{inputenc}
\usepackage[T1]{fontenc}

\usepackage{amstext}
\usepackage{amsmath}            %serve per le subequazioni
\usepackage{amssymb}            %serve per il simbolo "marchio registrato", \circledR

\newcommand{\etal}{\textit{et al.}}

\newcommand\xxx[1]{{#1}}

\newcommand\extra[1]{}
\begin{document}

\title{Entanglement-based linear-optical qubit amplifier}
\author{Evan Meyer-Scott}
\affiliation{Institute for Quantum Computing and Department of Physics and Astronomy, University of Waterloo, 200 University Avenue W, Waterloo, Ontario N2L 3G1, Canada }
\author{Marek Bula}
\affiliation{RCPTM, Joint Laboratory of Optics of Palacký University and Institute of Physics of Academy of Sciences of the Czech Republic, 17. listopadu 12, 771 46 Olomouc, Czech Republic}
\author{Karol Bartkiewicz} \email{bartkiewicz@jointlab.upol.cz}
\affiliation{RCPTM, Joint Laboratory of Optics of Palacký University and Institute of Physics of Academy of Sciences of the Czech Republic, 17. listopadu 12, 771 46 Olomouc, Czech Republic}
\author{Antonín Černoch}
\affiliation{Institute of Physics of Academy of Sciences of the Czech Republic, Joint Laboratory of Optics of PU and IP AS CR, 
   17. listopadu 50A, 772 07 Olomouc, Czech Republic}
\author{Jan Soubusta}
\affiliation{Institute of Physics of Academy of Sciences of the Czech Republic, Joint Laboratory of Optics of PU and IP AS CR, 
   17. listopadu 50A, 772 07 Olomouc, Czech Republic}
   \author{Thomas Jennewein}
   
\affiliation{Institute for Quantum Computing and Department of Physics and Astronomy, University of Waterloo, 200 University Avenue W, Waterloo, Ontario N2L 3G1, Canada }
\author{Karel Lemr}
\email{k.lemr@upol.cz}
\affiliation{Institute of Physics of Academy of Sciences of the Czech Republic, Joint Laboratory of Optics of PU and IP AS CR, 
   17. listopadu 50A, 772 07 Olomouc, Czech Republic}

\date{\today}

\begin{abstract}
We propose a linear-optical scheme for an efficient amplification of a photonic qubit based on interaction of the signal mode with a pair of entangled ancillae. In contrast to a previous proposal for qubit amplifier by Gisin \etal, [Phys Rev. Lett. {\bf 105}, 070501 (2010)] the success probability of our device does not decrease asymptotically to zero with increasing gain. Moreover we show how the device can be used to restore entanglement deteriorated by transmission over a lossy channel and calculate the secure key rate for device-independent quantum key distribution.
\end{abstract}


\maketitle

\section{Introduction}
\xxx{The fundamentals of quantum physics were discovered and formulated nearly a hundred years ago. 
Three decades ago scientists postulated that the laws of quantum physics could be used to 
improve capabilities of computation and communication technologies \cite{Wiesner83coding}.} This idea sparked intense research resulting in the discovery of many quantum information protocols, some of them even with practical, modern implementations  \cite{bib:nielsen:quantum_comput,bib:bruss:quantum_information}.

One such application of quantum information is quantum cryptography, comprising various quantum key distribution protocols (QKD) \cite{Gisin02crypto}. 
QKD offers unconditional security of private communications certified by the laws of quantum physics.
In the real world, QKD suffers from various technological limits, especially the need to trust imperfect detectors and single photon sources, quantum channel losses, and  
background noise.
The latter effects limit the maximum distance for unconditionally secure communications \cite{Bartkiewicz13}. 
Long-distance QKD has been realized over 144 km in free-space \cite{Ursin07QKD} 
and over 260 km in an optical fiber \cite{Wang12QKD-260km}. Trust in the imperfect devices used for cryptography allows eavesdroppers to attack unintended leakages of information or control detectors, known as side channels \cite{Makarov10Hacking}.

The side channel attacks can be solved in principle by using Bell-state projection measurements or using entanglement-based protocols. The simpler approach is measurement-device-independent QKD  
\cite{Lo12MDI-QKD,Chan12arXiv,Liu12arXiv}. 
In this case a projection on a 
Bell state in the middle of the communication line removes all detector side channels are removed.
The more complete approach is device-independent QKD (DI-QKD) 
\cite{gisin10ampl,acin07device,Kocsis13NPhys9,Pitkanen11ampl,Curty11ampl} 
and its security is based on 
the loophole-free violation of a Bell inequality. DI-QKD removes all source and detector side channels but requires closing of the detector (high-efficiency detection) and locality (distant detectors) loopholes, which has not yet been achieved simultaneously \cite{Zeilinger13Bell}. 

For DI-QKD and other protocols requiring high-efficiency detection, a method is required to circumvent the channel losses inherent in photon transmission. In classical optical communication networks the problem 
of losses is solved using amplifiers of the classical signal. For quantum communication, losses are more fundamental. The quantum signals are 
stored in polarization or temporal modes of individual photons and any quantum 
amplifier is bound by the quantum limits like the no-cloning theorem \cite{Wooters82Clone}. 
Several proposals of quantum amplifiers were recently introduced, wherein the quantum limit
can be circumvented by making the amplification non-deterministic. This type of 
amplification is called heralded noiseless amplification 
\cite{XiangNPhot4} and is already seeing successful implementation \cite{Kocsis13NPhys9}.
% %%%%%%%% 
%EMS: I commented out this section on alternative amplifiers because I found it distracting from our point, but please reinstate it if needed. 
%A completely different amplification approach is used in the case of 
%continuous-variable QKD where homodyne detection is used 
%\cite{Blandino12CV-QKD,Pooser09CV-amplif,Josse06amplif}.   
%Alternatively the problem of losses in QKD can be solved 
%performing a quantum nondemolition measurement of a photon's presence 
%\cite{Kok02QND,Bula13QKD}, or by using
%quantum repeaters \cite{Abruzzo13} to distribute entanglement between distant parties.
%%%%%%%%%%%
Note that there exists a complete equivalence between distribution of two-qubit
entanglement and secure key distribution \cite{acin03equiv}. In other words, 
any quantum channel is capable of secret communication if and only if 
it is capable of distributing entanglement.  

In this article we propose a new scheme of a linear-optical qubit amplifier
that can restore the attenuated qubit and is also capable of distilling deteriorated 
entanglement of the qubit state. Our amplifier is ready to be used in DI-QKD schemes. 
\xxx{Moreover it outperforms previously published proposals. 
In contrast to Gisin \etal\ scheme \cite{gisin10ampl}, the success probability of our 
device does not asymptotically approach zero when increasing the amplification gain.
Furthermore in comparison to Pitkanen \etal\ scheme \cite{Pitkanen11ampl}, our device 
provides tuneable gain and for the case of infinite gain allows better success probability due to its intrinsic elimination of the two-photon component after heralding. However, the Pitkanen {\em et al.} device may perform better when using a probabilistic source for the ancilla photons, due to its extra stage of heralding. The scheme by Curty and Moroder makes 
use of entanglement as in our device, but it is limited to infinite gain only \cite{Curty11ampl}, and in this regime it performs comparably to our device.
Further to these works, we present a thorough investigation of the
gain versus success probability tradeoff which is a crucial figure of merit for probabilistic amplifiers.  
}

The paper is organized as follows.
The principle of the amplifier operation is explained in sec.~\ref{sec_princip}.
The entanglement distillation is analyzed in sec.~\ref{sec_distill} and
DI-QKD is discussed in sec.~\ref{sec_DI-QKD}.
Conclusions are drawn in the final sec.~\ref{sec_conc}.

% -----------------------------------------------------------------------------

\section{Principle of operation\label{sec_princip}}
\begin{figure}
\includegraphics[scale=1]{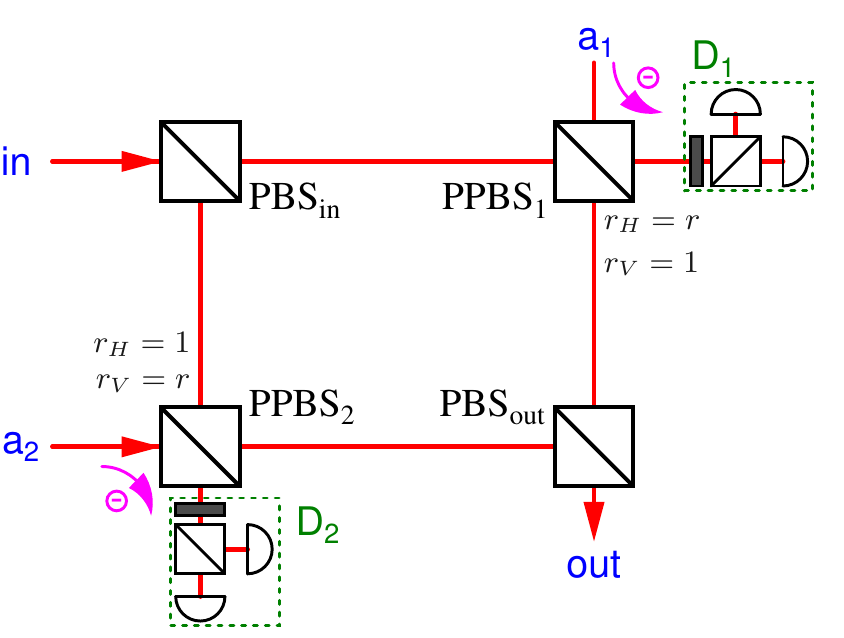}
\caption{\label{fig:scheme} (color online) Scheme for entanglement-based linear-optical qubit amplifier as described in the text. D$_1$ and D$_2$ are standard polarization analysis detection blocks (for reference see \cite{halenkova2012detector}).}
\end{figure}
The amplifier (depicted in Fig. \ref{fig:scheme}) consists of four polarizing beam splitters. Two of them (PBS$_\mathrm{in}$ and PBS$_\mathrm{out}$) form a Mach-Zehnder interferometer between signal input port ``in'' and output port ``out''. These polarizing beam splitters totally transmit horizontally polarized light while totally reflect light with vertical polarization. The other two are partially-polarizing beam splitters, denoted as PPBS$_1$ and PPBS$_2$, and placed in their respective arms of the interferometer. PPBS$_1$ reflects vertically polarized light, while having reflectivity $r$ for horizontal polarization. In terms of creation operators this transformation reads
\begin{eqnarray*}
\hat{a}^\dagger_\mathrm{in,H} & \rightarrow & r \hat{a}^\dagger_\mathrm{out,H} + \sqrt{1-r^2} \hat{a}^\dagger_\mathrm{D1,H}\\
\hat{a}^\dagger_\mathrm{a1,H} & \rightarrow & -r \hat{a}^\dagger_\mathrm{D1,H} + \sqrt{1-r^2} \hat{a}^\dagger_\mathrm{out,H}\\
\hat{a}^\dagger_\mathrm{a1,V} & \rightarrow & - \hat{a}^\dagger_\mathrm{D1,V},
\end{eqnarray*}
where labelling of spatial modes has been adopted from Fig. \ref{fig:scheme} and H, V denote horizontal and vertical polarizations. Similarly the PPBS$_2$ reflects completely the horizontal polarization and with reflectivity $r$ it reflects vertically polarized photons.
%\begin{eqnarray*}
%\hat{a}^\dagger_\mathrm{in,V} & \rightarrow & r \hat{a}^\dagger_\mathrm{out,V} + \sqrt{1-r^2} \hat{a}^\dagger_\mathrm{D2,V}\\
%\hat{a}^\dagger_\mathrm{a2,V} & \rightarrow & -r \hat{a}^\dagger_\mathrm{D2,V} + \sqrt{1-r^2} \hat{a}^\dagger_\mathrm{out,V}\\
%\hat{a}^\dagger_\mathrm{a2,H} & \rightarrow & - \hat{a}^\dagger_\mathrm{D2,H}.
%\end{eqnarray*}
The parameter $r$ is to be tuned as explained below. Successful operation of the amplifier is heralded by two-photon coincidence detection on detection blocks D$_1$ and D$_2$.

To demonstrate the principle of operation, let us assume the input signal to be a coherent superposition of vacuum and a polarization-encoded single photon qubit
$$
|\psi_\mathrm{in}\rangle = \alpha |0\rangle + \beta_H |H\rangle + \beta_V |V\rangle,
$$ 
where $|0\rangle$ denotes vacuum, $|H\rangle$, $|V\rangle$ denote horizontal and vertical polarization states respectively and the coefficients meet the normalization condition $|\alpha|^2+|\beta_H |^2+|\beta_V|^2=1 $. The amplifier makes also use of a pair of ancillary photons impinging on ports $a_1$ and $a_2$ of PPBS$_1$ and PPBS$_2$ respectively. These ancillary photons are initially in a maximally entangled Bell state of the form 
$$
|\Phi^+ _{a_1a_2}\rangle = \frac{1}{\sqrt{2}}(|H_{a_1}H_{a_2}\rangle + |V_{a_1}V_{a_2}\rangle),
$$
where the indices denote the ancillary photons' spatial modes.

The total state entering the amplifier composed of the signal and ancillary photons reads
\begin{eqnarray*}
|\psi_T\rangle &=& |\psi_\mathrm{in}\rangle \otimes |\Phi^+ _{a_1a_2}\rangle \\
&=&  \frac{1}{\sqrt{2}}\left[\alpha |0_\mathrm{in}H_{a1}H_{a2}\rangle + \alpha |0_\mathrm{in}V_{a1}V_{a2}\rangle\right.\\
 &+& \beta_H|H_\mathrm{in}H_{a1}H_{a2}\rangle+\beta_H|H_\mathrm{in}V_{a1}V_{a2}\rangle \\
&+& \left.\beta_V|V_\mathrm{in}H_{a1}H_{a2}\rangle+\beta_V|V_\mathrm{in}V_{a1}V_{a2}\rangle\right].
\end{eqnarray*}

Now we inspect evolution of all the individual terms present in previous equation.
Since the successful operation of the amplifier is conditioned by a two-photon coincidence detection by D$_1$ \& D$_2$ we post-select only such cases:
\begin{eqnarray*}
|0_\mathrm{in}H_{a1}H_{a2}\rangle &\rightarrow& r |0_\mathrm{out}H_{D1}H_{D2}\rangle \\
|0_\mathrm{in}V_{a1}V_{a2}\rangle &\rightarrow& r |0_\mathrm{out}V_{D1}V_{D2}\rangle \\
|H_\mathrm{in}H_{a1}H_{a2}\rangle &\rightarrow& (2r^2-1) |H_\mathrm{out}H_{D1}H_{D2}\rangle \\
|H_\mathrm{in}V_{a1}V_{a2}\rangle &\rightarrow& r^2 |H_\mathrm{out}V_{D1}V_{D2}\rangle \\
|V_\mathrm{in}H_{a1}H_{a2}\rangle &\rightarrow& r^2 |V_\mathrm{out}H_{D1}H_{D2}\rangle \\
|V_\mathrm{in}V_{a1}V_{a2}\rangle &\rightarrow& (2r^2-1) |V_\mathrm{out}V_{D1}V_{D2}\rangle .
\end{eqnarray*}

\xxx{Note that for $r = 0$, it is impossible to have more than one photon in the output mode, even for multiple photons in the input mode}. Subsequently we perform polarization-sensitive detection on D$_1$ and D$_2$ in the basis of diagonal $|D\rangle \propto (|H\rangle + |V\rangle)$ and anti-diagonal $|A\rangle \propto (|H\rangle - |V\rangle)$ linear polarization. This way we erase the information about the ancillary state and project the signal at the output port to
$$
|\psi_\mathrm{out}\rangle \propto \alpha r |0\rangle + \frac{3r^2-1}{2} \left(\beta_H |H\rangle + \beta_V |V\rangle\right),
$$   
where we have incorporated the fact that only if both the detected polarizations on D$_1$
and D$_2$ are identical (DD or AA coincidences) the device heralds a successful amplification and thus only one half of the measurement outcomes contributes to success probability.

At this point, we define the amplification gain $G$ as a fraction between signal and vacuum probabilities
\begin{equation}
\label{eq:gain}
G=\frac{(3r^2-1)^2}{4r^2} %% In PHYSICAL REVIEW A 86, 023815 (2012) and Nat Phys, 9, 23?28 (2013), Gain is defined as the single photon probability before and after amplification, including normalization.
\end{equation}
and calculate the corresponding success probability
\begin{equation}
\label{eq:psucc}
P=r^2\left[|\alpha|^2 + G \left(|\beta_H|^2+|\beta_V|^2\right)\right].
\end{equation}
Note that while the gain itself is input state independent, the success probability depends on both the gain and the input state parameters. This reflects the intuitive fact that it is for instance impossible to amplify a qubit that is actually not present in the input state ($\beta_H = \beta_V = 0$).

\begin{figure}
\includegraphics[scale=1]{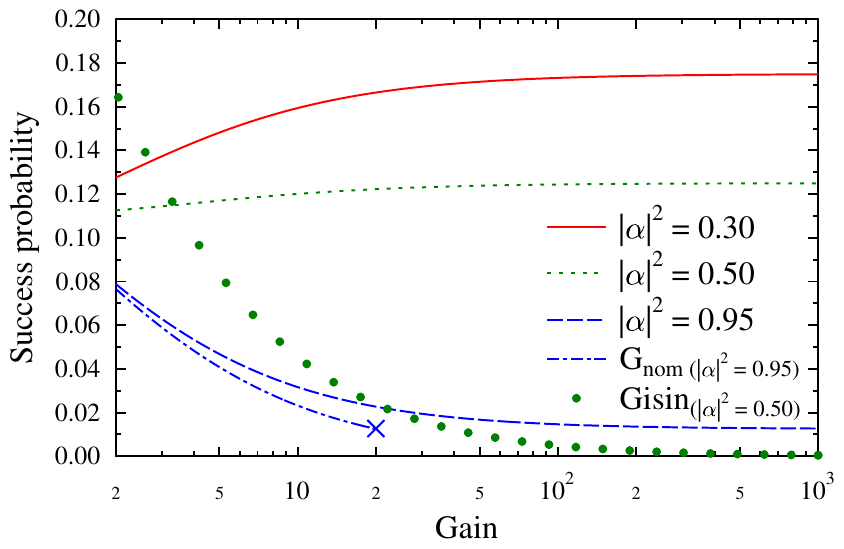}
\caption{\label{fig:prob_gain} (color online) Success probability is depicted as a function of gain for three different input states parametrized by $|\alpha|^2$. For comparison, the success probability of Gisin \etal~scheme \cite{gisin10ampl} is presented (in this case $|\alpha|^2 = 0.5$). Note that the success probability of our amplifier converges asymptotically to a non-zero value for any state with $|\alpha|^2\neq1$. Success probability is also plotted as a function of nominal gain $G_\mathrm{nom}$ for the case of $|\alpha|^2 = 0.95$. Note that according to its definition (\ref{eq:gain_nom}), the nominal gain is upper bounded by the value of 20 in this particular case (blue X symbol).}
\end{figure}
Let us analyse the results further. As expected the gain $G=1$ is obtained for $r=1$ with success probability $P=1$ independent on the input state. On the other hand, an infinite gain is obtained for $r=0$ with success probability of $P=(|\beta_H|^2+|\beta_V|^2)/4$. In this particular case, it is however possible to increase the success probability twice by including also detection coincidences DA and AD accompanied by a feed-forward operation $V\rightarrow-V$ on the output state. Note that this regime is suitable for non-demolition presence detection  of the qubit \cite{Bula13QND}. Fig. \ref{fig:prob_gain} depicts the trade-off between success probability and gain for three different input states containing different amounts of vacuum.

In a recent paper \cite{Kocsis13NPhys9}, its authors proposed also another measure of amplifier performance -- the nominal gain $G_\mathrm{nom}$ defined as
\begin{equation}
\label{eq:gain_nom}
G_\mathrm{nom} \equiv \frac{G}{|\alpha|^2 + G(|\beta_H|^2+|\beta_V|^2)} = \frac{r^2G}{P}.
\end{equation}
While the ordinary gain $G$ describes how much the qubit to vacuum intensity ratio has been increased under the amplification procedure, the nominal gain shows how much the overall success probability of finding the qubit state has increased. For this reason, the nominal gain is bound by the inverse value of the initial qubit probability (e.g. for $|\beta_H|^2 + |\beta_V|^2 = 0.2$, the maximum value of nominal gain is 5 and in this case the vacuum state is completely eliminated).  Fig. \ref{fig:prob_gain} depicts the success probability as a function of nominal gain for one particular initial state ($|\alpha|^2=0.95$).

It is worth noting that in contrast to Gisin \etal~scheme \cite{gisin10ampl}, the success probability does not decrease asymptotically to 0 with increasing gain (also illustrated in Fig. \ref{fig:prob_gain} for comparison). One may however suggest that in the case of infinite gain, the scheme performs exactly as well as standard teleportation. While this is indeed true, standard teleportation does not allow to tune the amplification gain and therefore the superposition of vacuum and qubit state collapses either onto vacuum or qubit state. In contrast, our scheme allows for the coherent superposition of these two terms to be maintained. Keeping coherence between vacuum and qubit terms is crucial for instance in all applications involving dual rail encoding.

\section{Amplification-based entanglement distillation\label{sec_distill}}

Quantum entanglement is one of the key ingredients in quantum communications. It can be used for teleportation \cite{bib:bouwmeester:teleport}, quantum cryptography \cite{bib:e91}, or remote state preparation \cite{barreiro2010remote}. It is also very sensitive to losses and decoherence occurring in the communication channel \cite{bartkiewicz07losses,halenkova2012noise,Horst13}. For this reason, entanglement distillation -- the way of improving entanglement of a state subjected to some degradation -- is a very important tool in quantum communications \cite{Yamamoto2001,Zhao2001}. In this section, we show how the amplifier can be used to distill entanglement on an example entangled state in dual-rail encoding.

Suppose an unknown polarization qubit $|\psi\rangle$ is distributed in two spatial modes creating thus maximally entangled state of the form 
\begin{equation}
\label{eq:entangled}
|\Psi\rangle = \frac{1}{\sqrt{2}}\left(|\psi 0\rangle + |0\psi\rangle\right).
\end{equation}
\xxx{States of vacuum and qubit superposition are needed in various quantum communication protocols
(e.g. quantum secret sharing \cite{Hillary99secret}) and are indispensable in implementations combining spatial and polarization encoding \cite{klmexper,router1,router2}.}
Now let us consider a lossy channel with transmissivity $1\geq T>0$ used to distribute the second spatial mode of this entangled state. This channel would deteriorate the state to 
$$
\hat{\rho}(\alpha,p) = (1-p)|00\rangle\langle 00| + p |\Psi_\alpha\rangle\langle\Psi_\alpha|,
$$
where 
$$|\Psi_\alpha\rangle = \sqrt{\alpha}|0\psi\rangle + \sqrt{1-\alpha}|\psi 0\rangle$$ 
with $\alpha = T/(T+1)$ and $p=(T+1)/2$. This state belongs to the class of amplitude damped states from Ref.~\cite{Horst13} where the entanglement and nonlocality of such states was studied. Since various measures of entanglement have different operational meaning, thus below we consider amplification of a few popular entanglement measures analyzed in \cite{Horst13} (for a review on entanglement measures see \cite{Horodecki09}). The negativity (concurrence) of the mixed state before amplification is simply $N=\frac{T}{2}$ ($C=\sqrt{T}$). After the amplification in the lossy mode the parameters of the state $\hat{\rho}(\alpha,p)$ read $\alpha = GT/(GT+1)$ and $p = \mathcal{N}(GT + 1)/2$ where $G$ denotes the gain as defined in previous section and $\mathcal{N} = 2/(2+GT-T)$. 
The entanglement of $\hat{\rho}(\alpha,p)$ (see Ref.~\cite{Horst13}) can be quantified by its concurrence 
$$
C = 2p\sqrt{\alpha(1-\alpha)} = \mathcal{N}\sqrt{GT}
$$
which can be further used to express its negativity as
\begin{eqnarray*}
N  &=&\frac{1}{2} [\sqrt{(1-p)^2 + C^2}  - (1-p)]\\ &=& \frac{\mathcal{N}}{2}[\sqrt{(1-T)^2 + 4GT } - (1-T)].
\end{eqnarray*}
The third prominent measure of entanglement is the relative entropy of entanglement $S$, but as demonstrated by Miranowicz and Ishizaka \cite{Miran08} finding closed formula for $S$ in case of the amplitude-damped states requires solving a single variable equation for which no general analytic solution is known. Hence, we calculate $S$ numerically as described in \cite{Miran08,Horst13}.

\begin{figure}
\includegraphics[scale=1]{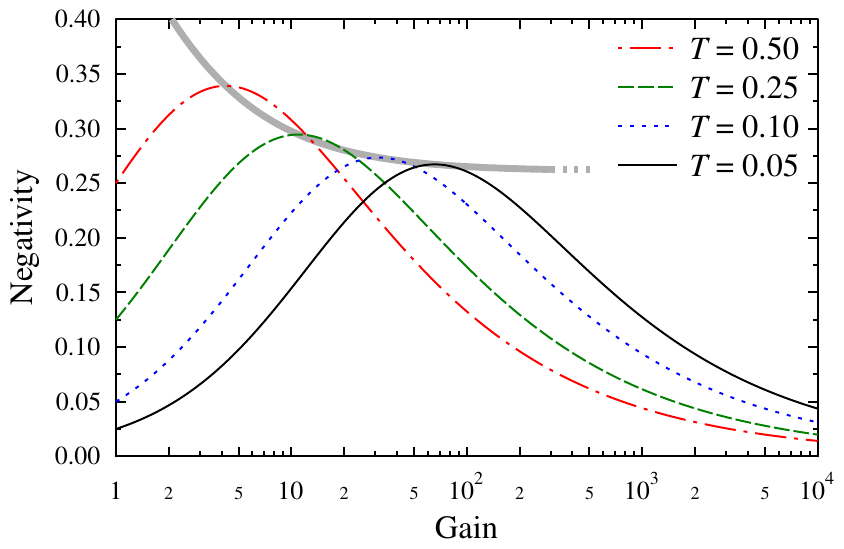}
\caption{\label{fig:ent_ampl} (color online) Negativity of entanglement depicted as a function of amplification gain for several different channel transmissivities $T$. A maximally entangled state formed of superposition of vacuum and qubit state is subjected to a channel with transmissivity $T$ resulting in entanglement loss. Suitably set amplification gain can increase the amount of entanglement. The wide grey curve joins the maxima of negativity for all values of transmissivity $T$ and subsequent optimal gains.}
\end{figure}

\begin{figure}
\includegraphics[scale=1]{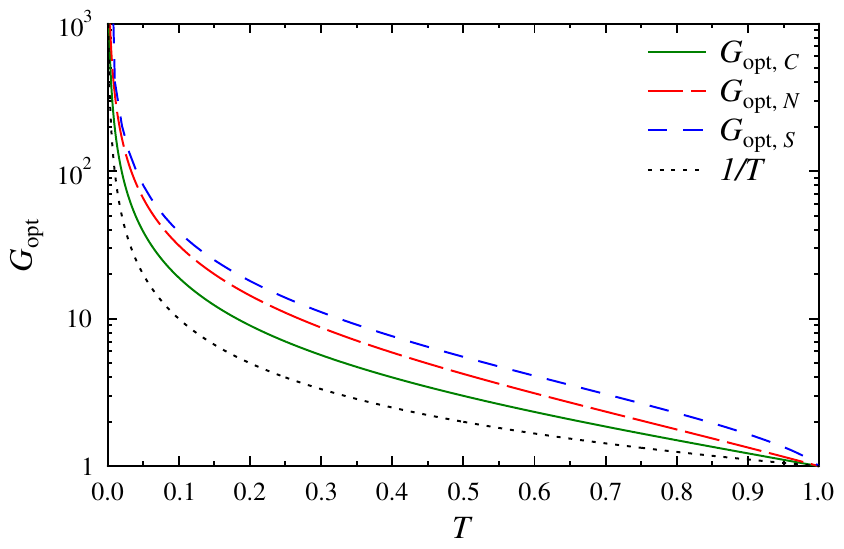}
\caption{\label{fig:opt_gain} (color online) The optimal gain $G_{\mathrm{opt}}$ for various entanglement measures as function of channel transmissivity $T$. Setting the optimal gain allows to obtain the largest possible value of the selected entanglement measure for a given loss parametrized by $T$. } 
\end{figure}

\begin{figure}
\includegraphics[scale=1]{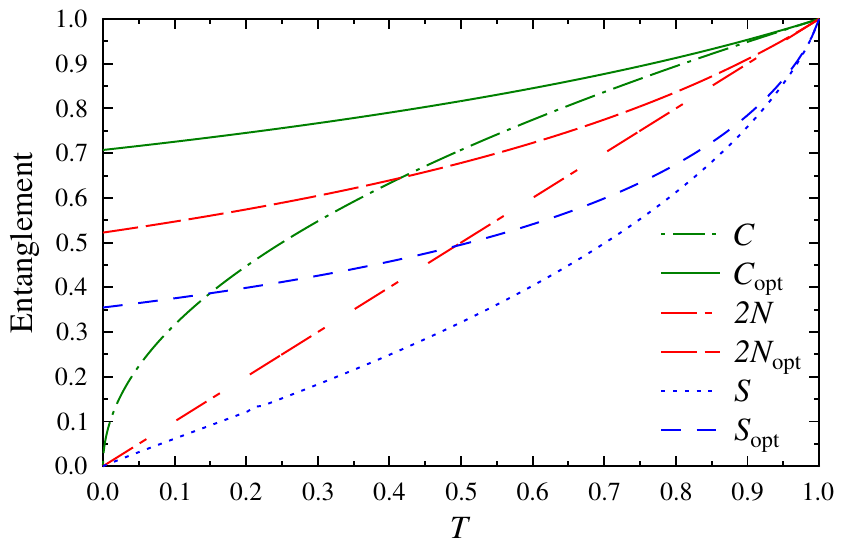}
\caption{\label{fig:opt_ent} (color online) The entanglement measures before ($C,\, N,\, S$) and after ($C_{\mathrm{opt}},\, N_{\mathrm{opt}},\, S_{\mathrm{opt}}$) optimal amplification   as functions of channel transmissivity $T$. } 
\end{figure}

As shown on the example of negativity in Fig.~\ref{fig:ent_ampl} the entanglement measures are functions both of transmissivity $T$ and gain $G$. The optimal gain for maximizing the entanglement is  
$$
G_{\mathrm{opt},N} = \frac{1}{T}[2-T-\sqrt{2-T} (T - 1)]
$$ 
for negativity and $ G_{\mathrm{opt},C} = (2-T)/T$ for concurrence. We do not present the exact expression for $S$ and its optimal gain, but the $G_{\mathrm{opt, S}}$ curve obtained numerically is presented together with other $G_{\mathrm{opt}}$ curves in Fig.~\ref{fig:opt_gain}. The curves shown in Fig.~\ref{fig:opt_gain} do not overlap, thus; the optimal gain $G_\mathrm{opt}$ varies depending on the entanglement  measure to be used. However, Fig.~\ref{fig:opt_gain} suggests that for any value of $T>0$, there is an optimal gain $G_{\mathrm{opt}} \geq \frac{1}{T}$ regardless of the applied entanglement measure. The entanglement measures before and after optimal amplification are depicted in Fig. \ref{fig:opt_ent} as functions of $T$. Note that for gain reaching infinity (standard teleportation), the entangled state would collapse onto the qubit state thus destroying the entanglement.

The corresponding success probability of the amplification process is
$$
P_\mathrm{succ} = \frac{r^2}{\mathcal{N}}= \frac{2G-2\sqrt{G^2 +3G} + 3}{9\mathcal{N}},
$$
where $r$ follows from Eq. (\ref{eq:gain}). In Fig. \ref{fig:ent_ampl} we plot the amplified negativity as a function of the chosen gain for several different values of channel transmissivity.  Note that our results for negativity, especially the expression for optimal gain $G_{\mathrm{opt},N}$, are also valid for logarithmic negativity $\log_2(2N+1)$ which is a concave function of $N$ providing an upper bound to the distillable entanglement \cite{Peres,Horodecki} given that the state was predistilled  using the above-described procedure. 

The above performed calculations reveal how qubit amplification can be used for partial entanglement recovery. However in neither of the cases, the entanglement has been restored to the original maximum value  due to the presence of the vacuum term $|00\rangle\langle 00|$. Recently, Mičuda \etal~experimentally demonstrated a rather clever way to eliminate the presence of such term \cite{micuda12ampl}. They considered only vacuum and a fixed polarization single photon state, but the technique can be adopted for qubit amplification as well. Their approach is based on deliberate coherent attenuation before the state is transmitted via the lossy channel. This coherent attenuation is performed by subjecting the state to a beam splitter with transmissivity $\nu$ and subsequent post-selection on vacuum in the ancillary mode. With the probability of $\nu$, one can thus disbalance the original state (\ref{eq:entangled}) to $|\Psi\rangle \to |\Psi_\alpha\rangle $
where $\alpha = \nu/(\nu+1)$. The choice of attenuation factor $\nu$ influences the probability $p=(1-\nu T)/(1+\nu)$ and $\alpha = \nu T/(1-\nu T) $ in the density matrix $\hat{\rho}(\alpha,p)$ of the state $|\Psi_\alpha\rangle$ transmitted through the lossy channel. Subsequent amplification will increase $\alpha$ thus also the entanglement of the state. Ideally for $\nu\rightarrow 0$ and gain $G\rightarrow \infty$ the original negativity can be completely restored. Of course such parameters lead to zero success rate so there is a need for some sort of compromise. Nevertheless this line of reasoning demonstrates the importance of amplification with high gain, where our amplifier outperforms the original Gisin \etal~proposal \cite{gisin10ampl}.

The above mentioned compromise can be quantified using the entangling efficiency $E_\mathrm{eff}$ of the protocol \cite{Lemr12}. The entangling efficiency is an entanglement generation measure suitable for probabilistic devices. In contrast to a more widely used entangling power \cite{ent-pow1,ent-pow2,ent-pow3}, the entangling efficiency optimizes over the device parameters in order to maximize  the product of success probability and negativity (or any other entanglement measure) 
$$
E_\mathrm{eff} = \mathrm{max}\lbrace P_\mathrm{succ}N\rbrace.
$$
\begin{figure}
\includegraphics[scale=1]{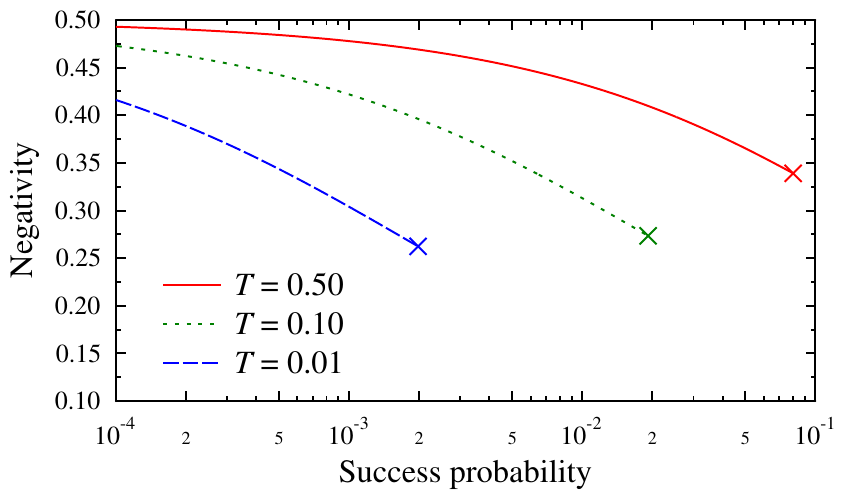}
\caption{\label{fig:prob_neg} (color online) Negativity and success probability trade-off obtained using coherent attenuation before transferring the state through a lossy channel. This trade-off is depicted for three different values of channel transmissivity $T$. Even though this strategy allows to increase the negativity arbitrarily close to $\frac{1}{2}$, the product of negativity and success probability is maximized when no coherent attenuation is used.}
\end{figure}
The negativity is calculated similarly as presented above using the analytical form of the density matrix. The success probability is composed of the success probability of attenuation ($\nu$) and the success probability of amplification (Eq. (\ref{eq:psucc})). In order to find the best strategy, we perform a numerical simulation. The plot in Fig. \ref{fig:prob_neg} shows the trade-off between negativity and success probability obtained when using the coherent attenuation  strategy. This simulation also reveals that the product of success probability and negativity is maximized for $\nu=1$ in all cases. So as far as the ``entanglement rate'' described by the entangling efficiency is concerned, the coherent attenuation does not offer any improvement. On the other hand, it is important to note that this strategy finds its merit when the goal is to achieve high negativity or high fidelity at the output.

\section{Device-independent quantum key distribution\label{sec_DI-QKD}}

Photon amplifiers can find additional applications in device-independent quantum key distribution, a stronger form of entanglement-based quantum cryptography based on the violation of Bell's inequality \cite{acin07device}. As mentioned above, DI-QKD does not require any knowledge of Alice and Bob's measurement devices, but does require closing the detection loophole \cite{pearle70hidden}. A number of ways of closing this loophole have been demonstrated, including using trapped ions \cite{rowe01exp,matsukevich08bell} and efficient photon detection \cite{giustina13bell}, but none has done so over the long distances needed for cryptography due to the intrinsic loss associated with photon transmission in fiber or free space. Gisin {\em et al.} recently proposed using a photon amplifier to herald incoming photons, closing the detection loophole and allowing DI-QKD \cite{gisin10ampl}. In their scheme, \xxx{as in the recently proposed improvements~\cite{Pitkanen11ampl,Curty11ampl},} a source of photons near Alice emits maximally entangled photon pairs. One photon is sent to Alice, which she detects directly with high efficiency, and the other photon is sent over a long channel to Bob. Bob routes the incoming photon through some heralded amplifier (e.g. the one proposed by Gisin {\em et al.} or by us) before detection, closing the detection loophole by performing a Bell measurement only upon successful amplification.

In order to compare the performance of the \xxx{three previous amplifiers} with ours, we performed numerical quantum optical simulations of the amplifiers. The initial source of entanglement was spontaneous-parametric down-conversion, with photon pair probability set to $2\times10^{-3}$, and both amplifiers used on-demand photon sources (two single photons for the Gisin {\em et al.} \xxx{and Pitkanen {\em et al.} schemes} and a maximally-entangled Bell state for ours) as ancillae. To mirror a likely experimental scenario, we used bucket detectors with 95~\% detection efficiency and 91~\% coupling efficiency as heralding detectors, and untrusted noiseless photon-number resolving detectors with the same efficiency for the detection of the photons for the Bell test after heralding. The former are modelled on fast superconducting nanowire detectors \cite{marsili13det} and the latter transition edge sensors \cite{lita08count}. We optimized all amplifiers over their tunable beam splitter reflectivity at each point. Finally we calculated the secure key rate per laser pulse from Eq.~(11) of the Supplementary Information of Ref. \cite{gisin10ampl}

\begin{equation}
R = \mu_{cc}\left[1 - h(Q) - I_E(S,\mu)\right],
\end{equation}
where $\mu_{cc}$ is the probability of a conclusive event for both Alice and Bob, $h(Q)$ is the binary entropy function of the measured quantum bit error rate, and $I_E(S,\mu)$ is Eve's information based on the Bell inequality violation $S$ and the ratio of inconclusive to conclusive results $\mu$ (see Eq.~(23) of Ref. \cite{gisin10ampl} for the full expression).

\begin{figure}
\includegraphics[scale=1]{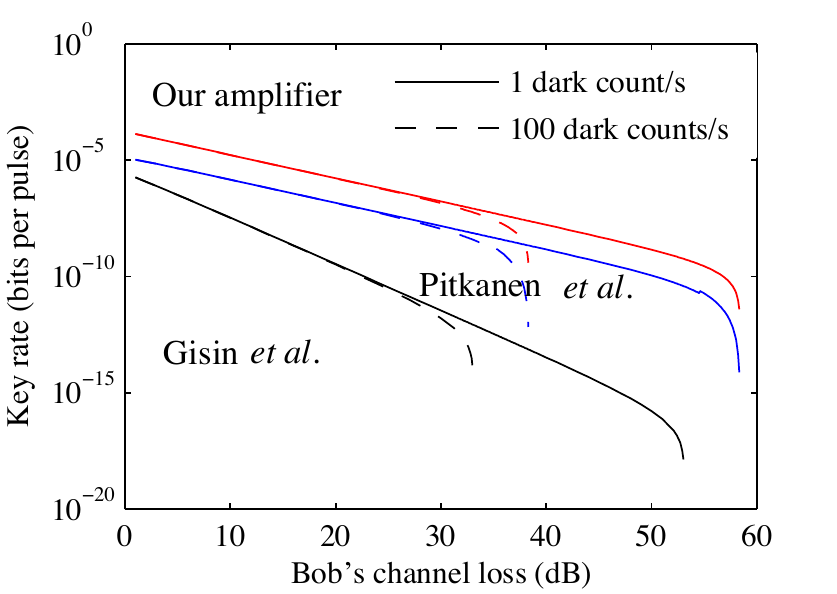}
\caption{\label{fig:keyrate} (color online) Key rate per laser pulse for device-independent quantum key distribution versus Bob's channel loss and dark counts per second in heralding detectors. Assuming 100 ps timing resolution in the heralding detectors leads to $10^{-10}$ and $10^{-8}$ dark count probability per pulse for 1 and 100 dark count/s respectively. Our entangled photon amplifier allows more key to be extracted than the Gisin {\em et al.} scheme, and even shows better scaling with loss. \xxx{It additionally delivers approximately 12 times the key rate of the Pitkanen {\em et al.} scheme.}}
\end{figure}

As shown in Fig. \ref{fig:keyrate}, our amplifier outperforms the Gisin {\em et al.} scheme and can also tolerate more dark counts in the heralding detectors.  This is because high gain is required to close the detection loophole after a lossy channel, and, as seen above, the success probability of the Gisin {\em et al.} photon amplifier converges asymptotically to zero for high gain. \xxx{It additionally outperforms the Pitkanen {\em et al.} scheme by a nearly constant factor, where this factor comes from improvements in success probability and the ratio of conclusive to inconclusive events after heralding. This is possible because in the Pitkanen {\em et al.} scheme, the elimination of the unwanted two-photon component even for ideal ancilla photons after heralding comes at the cost of vanishing success probability, a tradeoff our amplifier does not suffer from.} The optimal key rate in this \xxx{DI-QKD} scenario for our amplifier occurs with $r=0$ for all values of channel loss, \xxx{such that it performs identically to the Curty and Moroder proposal~\cite{Curty11ampl}.} However, there could be a regime (e.g. with noise in the final Bell test detectors) where higher success probability is needed to maximize key rate, at the cost of a larger vacuum component after the amplifier.

\section{Conclusion\label{sec_conc}}

In this paper, we have presented a linear-optical qubit amplifier. With the help of a maximally entangled photon pair, this device is able to change the ratio between vacuum and single qubit component, thus introducing qubit gain. In contrast to other proposals, our scheme achieves infinite gain with non-zero probability of success. Moreover, we have shown that the success probability of implementing infinite gain equals to the success probability of standard teleportation. To demonstrate the capabilities of our amplifier, we have presented two of its potential applications: entanglement distillation and quantum key distribution. Firstly, the analysis of entanglement distillation reveals that our amplifier can at least partially improve entanglement deteriorated by lossy transmission. We have presented the calculation of optimal gain for three different measures of entanglement (negativity, concurrence and relative entropy of entanglement) as a function of channel attenuation. Secondly, for device-independent quantum key distribution we have presented the significant improvement made by this amplifier over the previously proposed devices, including a key rate more than three orders of magnitude better for 100~km transmission distance.
\xxx{Practical implementation of the proposed scheme will be limited by available technology such
as precision of optical components, detection efficiency and delivery efficiency of ancillae.}

\begin{acknowledgments}
The authors gratefully acknowledge the support by the Operational Program Research and Development for Innovations -- European Regional Development Fund (project No. CZ.1.05/2.1.00/03.0058). A.~\v{C}. acknowledges project No. P205/12/0382 of Czech Science Foundation. K.~B. and K.~L. acknowledge support by Grant No. DEC-2011/03/B/ST2/01903 of the Polish National Science Centre and K.~B. also by the Operational Program Education for Competitiveness -- European Social Fund project No. CZ.1.07/2.3.00/30.0041. M.~B. acknowledges the financial support by Internal Grant Agency of Palack\'{y} University (No. PrF\_2013\_006). E.~M.~S. and T.~J. acknowledge support from the Natural Sciences and Engineering Research Council of Canada. \xxx{The authors thank Norbert L\"utkenhaus for helpful suggestions.}

\end{acknowledgments}

%Unused bibitems

%\bibitem{Clauser} J. F. Clauser, M. A. Horne, A. Shimony, and R. A. Holt,
%Phys. Lett. A \textbf{23}, 880 (1969).
%\bibitem{Bell}
%J. S. Bell, {Physics} (Lon Island City, N.Y.) \textbf{1}, 195
%(1964).
%\bibitem{kocsis13}
%S.~Kocsis, G.~Y.~Xiang, T.~C.~Ralph, and G.~J.~Pryde, Nat. Phys. {\bf 9}, 23--28 (2013)
\end{document}